\documentclass[]{IEEEphot}

\pdfoutput = 1

\usepackage{cite}
\usepackage{amsmath}

\usepackage{algorithm}  
\usepackage{algpseudocode}  

\usepackage{graphicx}
\usepackage{subfig}

\jvol{xx}
\jnum{xx}
\jmonth{June}
\pubyear{2009}

\begin{document}

\title{High-speed Implementation of FFT-based Privacy Amplification on FPGA in Quantum Key Distribution}

\author{Qiong Li, Bing-Ze Yan, Hao-Kun Mao, Xiao-Feng Xue}

\affil{Information Countermeasure Technique Institute, School of Computer Science and Technology,\\Harbin Institute of Technology, Harbin 150001 China. }  


\maketitle

\markboth{IEEE Photonics Journal}{High-speed Implementation of FFT-based Privacy Amplification on FPGA in QKD}

\begin{receivedinfo}%

\end{receivedinfo}

\begin{abstract}
Privacy amplification (PA) is a vital procedure in quantum key distribution (QKD) to generate the secret key that the eavesdropper has only negligible information from the identical correcting key for the communicating parties. With the increase of repeat frequency of discrete-variable QKD (DV-QKD) system, the processing speed of PA has become the bottle neck restricting DV-QKD's secure key rate. The PA using Toeplitz-based Hash function is adopted widely because of its simplicity and parallel feature. Because this algorithm can be accelerated with Fast Fourier Transform (FFT), an improved scheme PA for Field-programmable Gate Array (FPGA) based on this is proposed. This paper improves the custom FFT-based algorithm by reducing the number of computations and read/write memory operations significantly. The correctness is verified when implemented in a Xilinx Virtex-6 FPGA. Meanwhile, the processing speed of improved scheme can nearly double the classical Toeplitz Hashing scheme on FPGA through the actual experiment. 
\end{abstract}


\section{Introduction}

Quantum Key Distribution exploits quantum mechanics theorem to accomplish the secure key distribution. Since Bennet and Brassard proposed the first practicable protocol in 1984 \cite{Bennett2014f}, many protocols have been proposed successively. These protocols can divide into discrete variable (DV) protocols \cite{Stucki2005c,Scarani2004c,Inoue2002c,BruB1998c,Bennett1992e} and continuous variable (CV) protocols \cite{Cerf2001c,Grosshans2002f,Pirandola2008c,SUN2012c}. Because the DV-QKD is proposed earlier and the security proof of it is more complete, the development of DV-QKD drives to mature stage and many DV-QKD business systems have been developed \cite{Wang2012b,Muller2007b,Townsend1994b,Peng2007b}. We focused on the research of DV-QKD and found that key generation rate and QKD system on chip are two important research points of DV-QKD at this stage. DV-QKD is divided into four phases: quantum communication, public discussion, key reconciliation, and privacy amplification. The first three phases make two distant legitimate parties, usually named Alice and Bob, obtain identical random sequence. However, in this process, some information is exposed inevitably to Eavesdropper, usually named Eve. 

Privacy amplification(PA) eliminates the leaked information by distilling the final secret key from a long secret random sequence with universal Hash function \cite{Bennett1995d,Bennett1988d,Impagliazzo1989c}. Several classes of Hash function has been applied to perform the PA \cite{Wegman1981b,Carter1979b}. To reduce the finite size effect in distilling secure keys, the lengths of input blocks for PA should be at least $10^6$\cite{Cai2009c}, and this leads to large length of processing blocks. Zhang et al. choose a simple multiplicative universal class of Hash function to speed up PA process, and they construct an optimal multiplication algorithm with four basic multiplication algorithms\cite{Zhang2014c}. The speed of this algorithm achieves 14.68 Mbps based on CPU, but this algorithm is an iterative algorithm meaning that it consumes large resources and it is unsuited for hardware implementation. Besides, Toeplitz hashing\cite{Krawczyk1994b} is widely used in PA process because of its simplicity and parallel feature. The authors in \cite{Zhang2012b} proposed block parallel algorithm to speed Toeplitz hashing. The authors in \cite{Constantin2017e,Yang2017c} proposed improved block parallel algorithm of Toeplitz hashing respectively. The algorithm in \cite{Yang2017c} achieves 64 Mbps processing speed based on field-programmable gate array (FPGA) and reduces memory resources significantly. While this kind of algorithm has reach its speed limit constrained of its algorithmic complexity $O(n^2)$. Fast Fourier transform (FFT) and fast number theory transform (FNTT) are efficient fast algorithms for Toeplitz hashing, which reduce the algorithmic complexity from $O(n^2)$ to $O(n\log n)$. The authors in \cite{Liu2016d} firstly proposed a FFT based PA algorithm and implemented on Many Integrated Core (MIC). The process speed of algorithm reaches 60Mbps with raw key length of 12.8M. The authors in \cite{XiangyuWangYichenZhangSongYu2016e} proposed a FFT-based PA algorithm in CV-QKD based on graphic processing unit (GPU). The speed of privacy amplification is achieved over 1 Gbps. However, this algorithm is only suitable for CV-QKD, because it's efficient in case of great raw key length and low compression ratio. Crucially, the GPU and MIC platform are both hard to be integrated for its volume and power consumption.

Based on the investigation result, FPGA is a suitable platform for DV-QKD system with the feature of high-parallelism and embeddable platform. More importantly, the energy consumption of FPGA is much lower than that of GPU and MIC. Nevertheless, existing PA algorithms on FPGA are all parallel block method of Toeplitz hashing with algorithmic complexity of $O(n^2)$. FFT have great potential to accelerate the speed and reduce the consumption of PA algorithm on FPGA. Unfortunately, there is no practicable FFT-based PA scheme on FPGA. The main reason is that the $10^6$ length of input blocks increases the difficulty and cost of implementation. To solve this problem, we proposed a FFT-based PA hardware accelerate algorithm and implemented it on Virtex-6 FPGA. The throughput of our algorithm reaches 116Mbps with the raw key length $n=1M$. It's nearly 2 times faster than the classical Toeplitz Hashing Algorithm on FPGA.

The rest of this paper is organized as follows. The principle of privacy amplification with modified Toeplitz we used is described in Section 2. In Section 3, the detail and the key improvements of our FFT-based PA hardware accelerate algorithm are stated. In Section 4, we present our PA hardware implementation module and experiment results, including the processing speed and the requirements of the hardware resources. A comparison between several FPGA-based hardware implementations is also presented. In Section 5, we provide a brief conclusion.

\section{Related Work}

\subsection{Privacy Amplification}
\begin{figure}[!htbp]
	\centering
	\includegraphics[width=15pc]{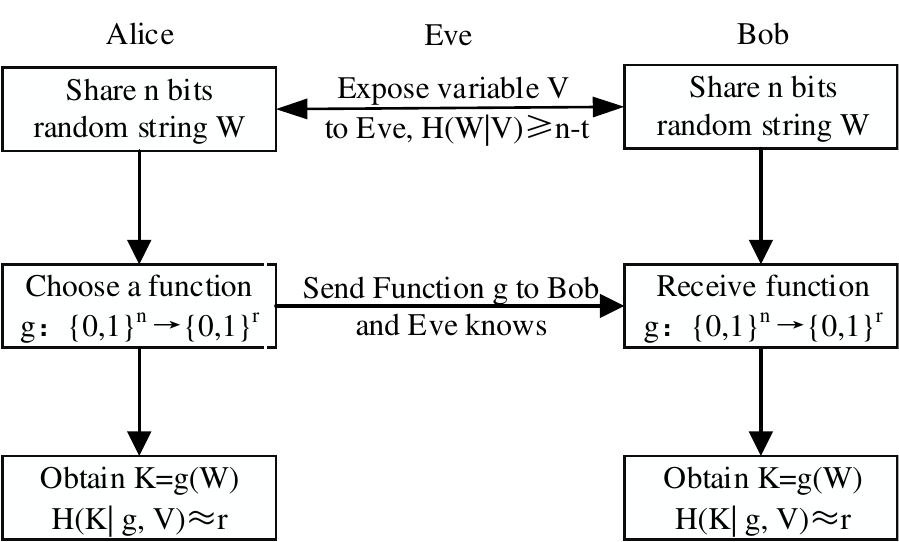}
	\caption{The procedure of privacy amplification}
	\label{fig_PA}
\end{figure}

Privacy amplification is a process that allows two parties to distill a secret key from a secret random variable about which an eavesdropper has partial information\cite{Bennett1995d}. Before PA procedure in QKD, Alice and Bob share a random $n$ bits binary string $W$, while Eve learns a correlated random string $V$ providing $t$ $(t<n)$ bits of information about $W$, i.e., $H(W|V)\ge n-t$. Alice and Bob wish to publicly choose a compression function $g:\{0,1\}^n \to \{0,1\}^r$ such that Eve's partial information on $W$ and her complete information on $g$ give her arbitrarily little information about $K=g(W)$ . this procedure is indicated as Fig.~\ref{fig_PA}.

Universal hash function\cite{Krawczyk1994b} is discovered to suit as the compression function g for privacy amplification \cite{Bennett1995d}. The mutual information between the distill key compressed by universal hash function and Eve's information can be proved using R\'enyi entropy,
\begin{equation}
I(K;g,V)\le {{2}^{-s}}/\ln 2
\label{eq1}
\end{equation}%
Where $s=n-t-r$ means the security coefficient of PA.

Considering the security threaten of finite-key effect, the input raw key length n of privacy amplification in DV-QKD should be larger than $10^6$\cite{Cai2009c,Scarani2008c}. Therefore, the calculation of hash function is very large, and the choice of hash function class is extremely important.
\subsection{Modified Toeplitz Matrix}
The Toeplitz matrix is a particular class of the universal hash functions\cite{Krawczyk1994b}. Because Toeplitz matrix is diagonal-constant matrix, Toeplitz matrix can be constructed by its first column and first row and calculated by FFT. Therefore, the required number of random bits can be reduced to $n+r-1$ and the calculation complexity can be reduced to $O(n\log n)$. However, the input length of FFT would be $n+r-1$ or $2n$ to calculate the Toeplitz matrix depending on the compression radio, that would cost a lot of extra cost in hardware implementation.

Hayashi et al. proposed using modified Toeplitz matrix instead of Toeplitz matrix as the compression function and give the security proof \cite{Urumaru2013a}. The modified Toeplitz matrix is constructed by the concatenation of Toeplitz matrix and the identity matrix $(X,I)$. For instance, Eq.\eqref{eq2} is a modified Toeplitz matrix.

\begin{equation}
{{\mathbf{S}}_{\mathbf{r}\times \mathbf{n}}}=\left[ \begin{matrix}
1 & {} & {} & {} & {} & {{V}_{r-1}} & {{V}_{r-2}} & \cdots  & {{V}_{n-2}} & {{V}_{n-1}}  \\
{} & 1 & {} & {} & {} & {{V}_{r-2}} & {{V}_{r-1}} & {} & {} & {{V}_{n-2}}  \\
{} & {} & 1 & {} & {} & \vdots  & {} & \ddots  & {} & \vdots   \\
{} & {} & {} & 1 & {} & {{V}_{1}} & {} & {} & {{V}_{n-r-1}} & {{V}_{n-r}}  \\
{} & {} & {} & {} & 1 & {{V}_{0}} & {{V}_{1}} & \cdots  & {{V}_{n-r-2}} & {{V}_{n-r-1}}  \\
\end{matrix} \right]
\label{eq2}
\end{equation}%

Using the modified Toeplitz matrix, the required quantity of random bits can be reduced to n and the input length of FFT can be reduced to n. In this case, the length of FFT is only related to the input block length of the key other than the final key length. This feature notably reduced the design complexity of FFT-based PA.

\subsection{Modified  Toeplitz Matrix Calculation by FFT}

FFT algorithm is a common algorithm to calculate the Toeplitz matrix due to the  $O(n\log n)$ calculation complexity of FFT. A general calculative process of the modified Toeplitz is provided in this section. ${{\mathbf{X}}_{\mathbf{n}}}={{\left[ {{X}_{0}},{{X}_{1}},\cdots ,{{X}_{n\text{-}1}} \right]}^{'}}$ is the input sequence of PA. ${{\mathbf{Y}}_{\mathbf{r}}}={{\left[ {{Y}_{0}},{{Y}_{1}},\cdots ,{{Y}_{r}} \right]}^{'}}$ is the final key sequence. Eq.\eqref{eq3} is the calculative process of the final key with the modified Toeplitz matrix $S_{r\times n}$.
\begin{equation}
{{\mathbf{Y}}_{\mathbf{r}}}={{\mathbf{S}}_{\mathbf{r}\times \mathbf{n}}}\times {{\mathbf{X}}_{\mathbf{n}}}=[{{\mathbf{I}}_{\mathbf{r}\times \mathbf{r}}},{{\mathbf{V}}_{\mathbf{r}\times \mathbf{(n-r)}}}]\times [\begin{matrix}
{{\mathbf{X}}_{\mathbf{r}}}  \\
{{\mathbf{X}}_{\mathbf{n-r}}}  \\
\end{matrix}]=[{{\mathbf{I}}_{\mathbf{r}\times \mathbf{r}}}\times {{\mathbf{X}}_{\mathbf{r}}}]+[{{\mathbf{V}}_{\mathbf{r}\times \mathbf{(n-r)}}}\times {{\mathbf{X}}_{\mathbf{n-r}}}]\text{=}{{\mathbf{X}}_{\mathbf{r}}}+\mathbf{Y}_{\mathbf{r}}^{\mathbf{'}}
\label{eq3}
\end{equation}%

Make up the Toeplitz matrix $V_{r\times (n-r)}$  to the cyclic matrix, then the calculation of $Y^{'}_r$  can be accelerated by FFT shown as Eq.\eqref{eq4}.

\begin{equation} 
\begin{split}
&[\begin{matrix}
{{\mathbf{P}}_{\mathbf{(n-r)}\times \mathbf{(n-r)}}}\times {{\mathbf{X}}_{\mathbf{n-r}}}  \\
\mathbf{Y}_{\mathbf{r}}^{\mathbf{'}}  \\
\end{matrix}]\text{=}[\begin{matrix}
{{\mathbf{P}}_{\mathbf{(n-r)}\times \mathbf{(n-r)}}}\times {{\mathbf{X}}_{\mathbf{n-r}}}  \\
{{\mathbf{V}}_{\mathbf{r}\times \mathbf{(n-r)}}}\times {{\mathbf{X}}_{\mathbf{n-r}}}  \\
\end{matrix}]\text{=}\left[ \begin{matrix}
{{\mathbf{P}}_{\mathbf{(n-r)}\times \mathbf{r}}} & {{\mathbf{P}}_{\mathbf{(n-r)}\times \mathbf{(n-r)}}}  \\
{{\mathbf{P}}_{\mathbf{r}\times \mathbf{r}}} & {{\mathbf{V}}_{\mathbf{r}\times \mathbf{(n-r)}}}  \\
\end{matrix} \right]\times [\begin{matrix}
\mathbf{0}  \\
{{\mathbf{X}}_{\mathbf{n-r}}}  \\
\end{matrix}] \\
&\qquad\qquad\qquad\qquad\quad\:\:\:={{\mathbf{V}}_{\mathbf{n}}}\otimes \mathbf{X}_{\mathbf{n}}^{\mathbf{'}}=\operatorname{IFFT}(\operatorname{FFT}({{\mathbf{V}}_{\mathbf{n}}})\bullet \operatorname{FFT}(\mathbf{X}_{\mathbf{n}}^{\mathbf{'}})) \\
\end{split}
\label{eq4}
\end{equation}%

 ${{\mathbf{V}}_{\mathbf{n}}}=[{{V}_{0}},{{V}_{1}},\cdots ,{{V}_{n}}]$ is the description of Toeplitz matrix,  $\mathbf{X}_{\mathbf{n}}^{\mathbf{'}}={{\left[ 0,0,\cdots ,{{X}_{n-r}},\cdots ,{{X}_{n-1}} \right]}^{'}}$ is a part of the input sequence. The matrix $P$ aims to complement the Toeplitz matrix ${{\mathbf{V}}_{\mathbf{r}\times \mathbf{(n-r)}}}$ to cyclic matrix.
 
\section{High Speed FFT-based Privacy Amplification Hardware Scheme }

A high speed PA hardware scheme for FPGA implementation is proposed in this section. An overall process of the scheme is given based on FFT algorithm. Furthermore, three points of Algorithm optimization are given in accordance with the feature of privacy amplification.

\subsection{Overall Process of FFT-based PA Hardware Schemes}

\begin{figure}[!h]
	\centering
	\includegraphics[width=25pc]{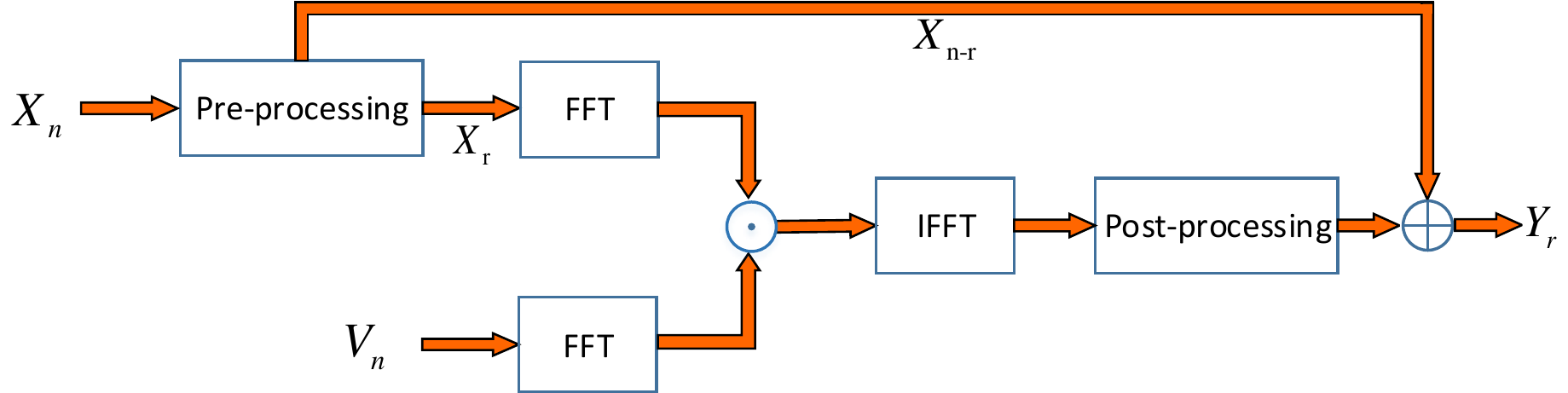}
	\caption{The Overall Process of FFT-based PA Hardware Scheme}
	\label{fig_FFT}
\end{figure}

The overall process of FFT-based PA Hardware Scheme is indicated as Fig.\ref{fig_FFT}. The pre-processing phase divides the input sequence ${{\mathbf{X}}_{n}}={{\left[ {{X}_{0}},{{X}_{1}},\cdots ,{{X}_{n\text{-}1}} \right]}^{'}}$ into ${{\mathbf{X}}_{r}}={{\left[ {{X}_{0}},{{X}_{1}},\cdots ,{{X}_{r\text{-}1}},0,\cdots ,0 \right]}^{'}}$ and ${{\mathbf{X}}_{n-r}}={{\left[ 0,0,\cdots ,0,{{X}_{r}},\cdots ,{{X}_{n-1}} \right]}^{'}}$ . The dot operational character means the dot product of the two FFT results. The post-processing phase rounds the result of IFFT to Boolean sequence. The final key sequence of PA is the XOR result of ${{\mathbf{X}}_{n-r}}$ and the result of the Toeplitz cyclic convolution.

The calculation of FFT/IFFT needs the greatest computation in the entire process. Although special hardware circuits for FFT/IFFT is provided in FPGA, the input length of these circuits can’t satisfy the request of PA. Concerning this issue, two-dimensional long FFT algorithm\cite{LernerParallel2018Spec} is adopted to calculate long FFT with small point FFT core. the procedure of custom 2-D long FFT algorithm is described as Algorithm \ref{algorithm_custom} .

In this way the calculation of FFT and IFFT can be accomplished with multiple small point FFT cores at high speed. However, this algorithm needs matrix transposition and storage repeatedly, that leads to large amount of memory read and write operations. Thus, the data transfer rate of memory is the bottleneck of the entire PA algorithm. Aiming at such shortcomings, several optimizations are proposed to reduce the amount of storage data and improve the processing speed according to the feature of PA.

\subsection{Real-valued FFT Acceleration}

The input sequence $X_n$ and the description of Toeplitz $V_n$ are both real sequences, but the FFT hardware circuits are designed to compute the FFT of a complex sequence. Most  FFT-based PA scheme regards the input sequence as the real part and sets imaginary part to 0 directly . This method leads to a waste of computing resource and storage resource. A real-valued FFT algorithm\cite{Sorensen1987a} is introduced in our method to solve this problem. With this algorithm, two real-valued FFT calculations can be accomplished by one complex-valued FFT. In our scheme, we need the FFT results of the input sequence and the Toeplitz sequence. We can get the results in one complex-valued FFT with the method described as below.
\begin{equation}
z(n)=x(n)+i\cdot v(n)
\label{eq5}
\end{equation}
\begin{equation}
Z(k)=FFT(z(n))
\label{eq6}
\end{equation}
\begin{equation}
\operatorname{Re}[X(k)]=1/2\cdot(\operatorname{Re}[Z(k)]+\operatorname{Re}[Z(N-k)])
\label{eq7}
\end{equation}
\begin{equation}
\operatorname{Im}[X(k)]=1/2\cdot(\operatorname{Im}[Z(k)]-\operatorname{Im}[Z(N-k)])
\label{eq8}
\end{equation}
\begin{equation}
\operatorname{Re}[V(k)]=1/2\cdot(\operatorname{Im}[Z(k)]+\operatorname{Im}[Z(N-k)])
\label{eq9}
\end{equation}
\begin{equation}
\operatorname{Im}[V(k)]=1/2\cdot(\operatorname{Re}[Z(N-k)]-\operatorname{Re}[Z(k)])
\label{eq10}
\end{equation}
This optimization accomplishes the FFT of two sequence with one complex-valued FFT operation, improving the processing rate and saving nearly half of the computing resource and storage resource.

\begin{algorithm}[hbt]
	\caption{Custom 2-Dimensional Long FFT of x}
	\begin{algorithmic}[1]
		\Require $X_{n}=x_{0},x_{1},\cdots,x_{n-1}$
		\Ensure  $Y_{n}=FFT(X_{n})$
		\State   Convert one-dimensional input sequence $X_{n}$ into two-dimensional matrix $A_{k\times k}$
		\State 	 $A^{'} =Transposed(A)$  \qquad \qquad \qquad \    // Transposed(A) is the transpose function of the matrix A
		\For     {$i=0$ to $k-1$}
		\State   $A_1[i][0:k-1] = FFT(A^{'}[i][0:k-1])$
		\EndFor
		\For {$i=0$ to $k-1$}
		\For {$j=0$ to $k-1$}
		\State $A_{2}[i][j] = A_{1}[i][j]\times W[i\times j]$   //W is the multiply rotation factor 
		\EndFor
		\EndFor
		\State 	 $A^{'}_{2} =Transposed(A_2)$
		\For {$i=0$ to $k-1$}
		\State  $A_3[i][0:k-1] = FFT(A^{'}_{2}[i][0:k-1])$
		\EndFor
		\State $A^{'}_{3} =Transposed(A_3)$
		\For {$i=0$ to $k-1$}
		\For {$j=0$ to $k-1$}
		\State	$Y[i\times k+ j] = A^{'}_{3}[i][j]$
		\EndFor
		\EndFor
	\end{algorithmic}
	\label{algorithm_custom}
\end{algorithm}

\subsection{A Modified 2D-FFT for PA}

The matrix transposition and storage operation of long FFT algorithm will cost an amount of time. Nevertheless, in the PA, the unconditional secure key is the final result of PA and the security of the key is not affected by the input sequence of the origin key and the export order of the final key. Therefore, the input and output order of FFT algorithm result is also out of consideration in PA algorithm. Without regard to the order of FFT algorithm, the procedure of long FFT/IFFT algorithm mentioned earlier (3.1) can be simplified as follow:

\begin{algorithm}[htb]
	\caption{Modified 2-D FFT for PA }
	\begin{algorithmic}[1]
		\Require $X_{n}=x_{0},x_{1},\cdots,x_{n-1}$
		\Ensure  $Y_{n}=FFT(X_{n})$
		\State   Convert one-dimensional input sequence $X_{n}$ into two-dimensional matrix $A_{k\times k}$            
		\For     {$i=0$ to $k-1$}
		\State   $A_1[i][0:k-1] = FFT(A[i][0:k-1])$
		\EndFor
		\For {$i=0$ to $k-1$}
		\For {$j=0$ to $k-1$}
		\State $A_{2}[i][j] = A_{1}[i][j]\times W[i\times j]$   //W is the multiply rotation factor 
		\EndFor
		\EndFor
		\State 	 $A^{'}_{2} =Transposed(A_2)$// Transposed(A) is the transpose function of the matrix A
		\For {$i=0$ to $k-1$}
		\State  $A_3[i][0:k-1] = FFT(A^{'}_{2}[i][0:k-1])$
		\EndFor
		\For {$i=0$ to $k-1$}
		\For {$j=0$ to $k-1$}
		\State	$Y[i\times k+ j] = A_{3}[i][j]$
		\EndFor
		\EndFor
	\end{algorithmic}
\end{algorithm}

\begin{figure}[!htp]
	\centering
	\begin{minipage}[t]{0.4\linewidth}
	\centering
	\includegraphics[width=6cm,height=6cm]{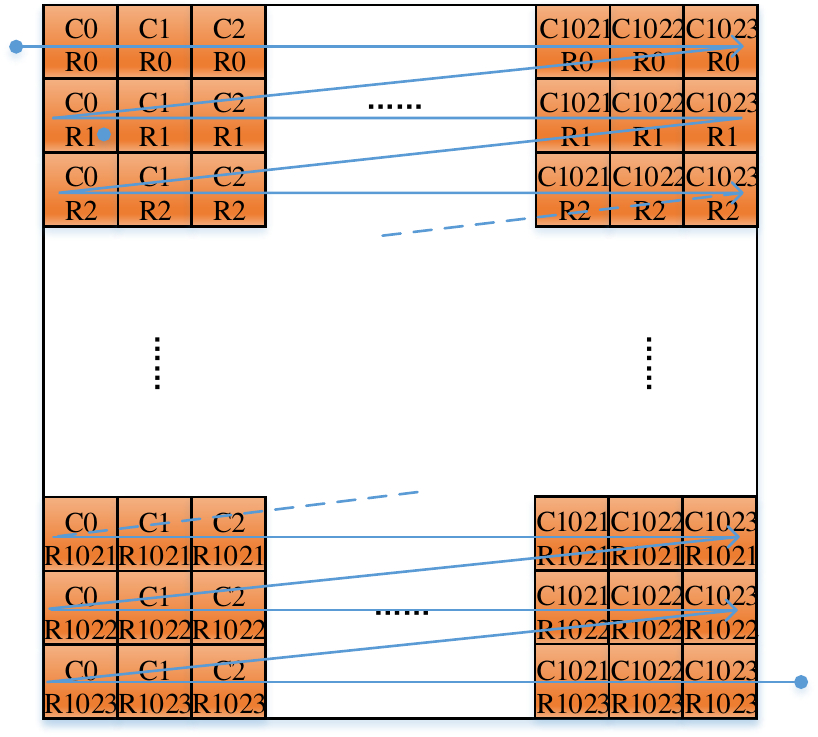}
	\caption*{(a)Custom 2-D FFT}
	\label{fig3:side:a}
	\end{minipage}
	\begin{minipage}[t]{0.4\linewidth}
	\centering
	\includegraphics[width=6cm,height=6cm]{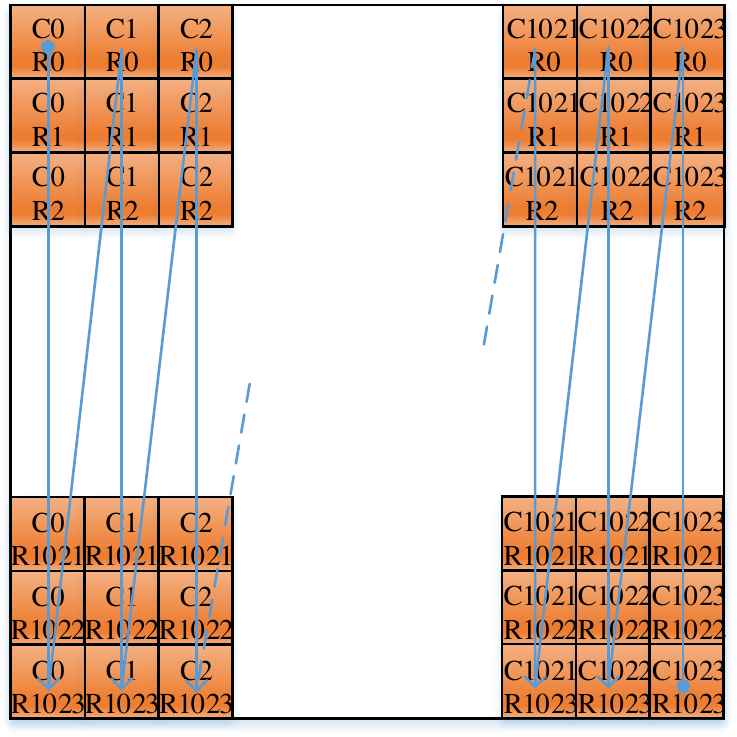}
	\caption*{(b) Modified 2-D FFT for PA}
	\label{fig3:side:b}
	\end{minipage}
	\caption{the Input/Output Sequence Order Diagram of FFT}
	\label{fig3}
\end{figure}

In this way the matrix transformation and storage that the long FFT/IFFT algorithm needs will significantly decrease. Taking the one-million points PA algorithm as an example, the output sequence of the PA algorithm with the natural order FFT and the unnatural order FFT is indicated as fig.\ref{fig3}.

Because our one-million points PA algorithm is based on the two-dimensional FFT algorithm, the input sequence will be loaded into a 1024-demensional matrix. If the matrix is processed row by row in the two-dimensional FFT algorithm, the input and output sequence of the PA algorithm with the natural order FFT is identical to that in Fig. \ref{fig3} (a). Meanwhile, the input and output sequence of the PA algorithm with the unnatural order FFT is shown in Fig. \ref{fig3} (b) in the column-by-column mode. The unnatural order optimizing changes the sequence of the final key and doesn’t affect the security of the final key. The times of matrix transposition and storage operation is decreased from six times to twice.

\subsection{Fast Matrix Transposition}

\begin{figure}[htp]
	\centering
	\includegraphics[width=8cm]{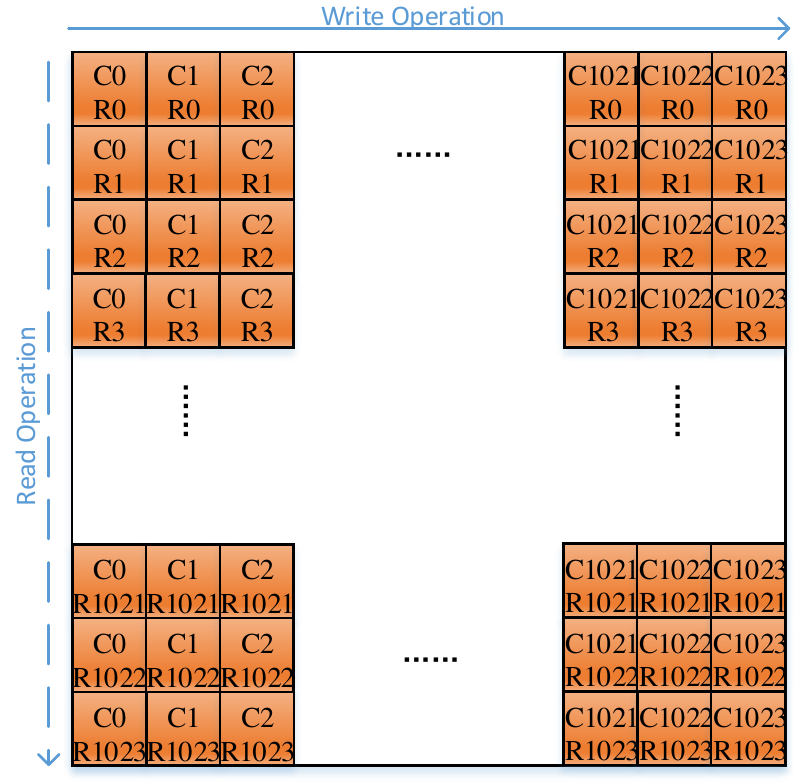}
	\caption{ the Common Matrix Transposition Process}
	\label{fig_Com_Matrix}
\end{figure}

Although modified 2D-FFT algorithm has decreased the matrix transposition times to twice, the matrix transposition still spends a lot of time. A high effective matrix transposition method\cite{Qinwen2017} is introduced in our scheme. Due to the access mechanism of DDR-SDRAM, the row span access operation will reduce the access speed of DDR. However, the matrix transposition needs a large amount of the row span access operations. The high effective matrix transposition method uses the matrix partitioning technology to reduce the times of the row span access operation. Taking the one-million points PA algorithm as an example, the main process of the common matrix transposition based on the DDR memory model is indicated as Fig. \ref{fig_Com_Matrix}.\\

The row span access operation times of the common matrix transposition is calculated as Eq. \ref{eq11}:

\begin{equation}
{{T}_{row-span}}={{T}_{write}}+{{T}_{read}}=1024+1024\times 1024\text{= }1049600
\label{eq11}
\end{equation}

\begin{figure}[h]
	\centering
	\includegraphics[width=14cm]{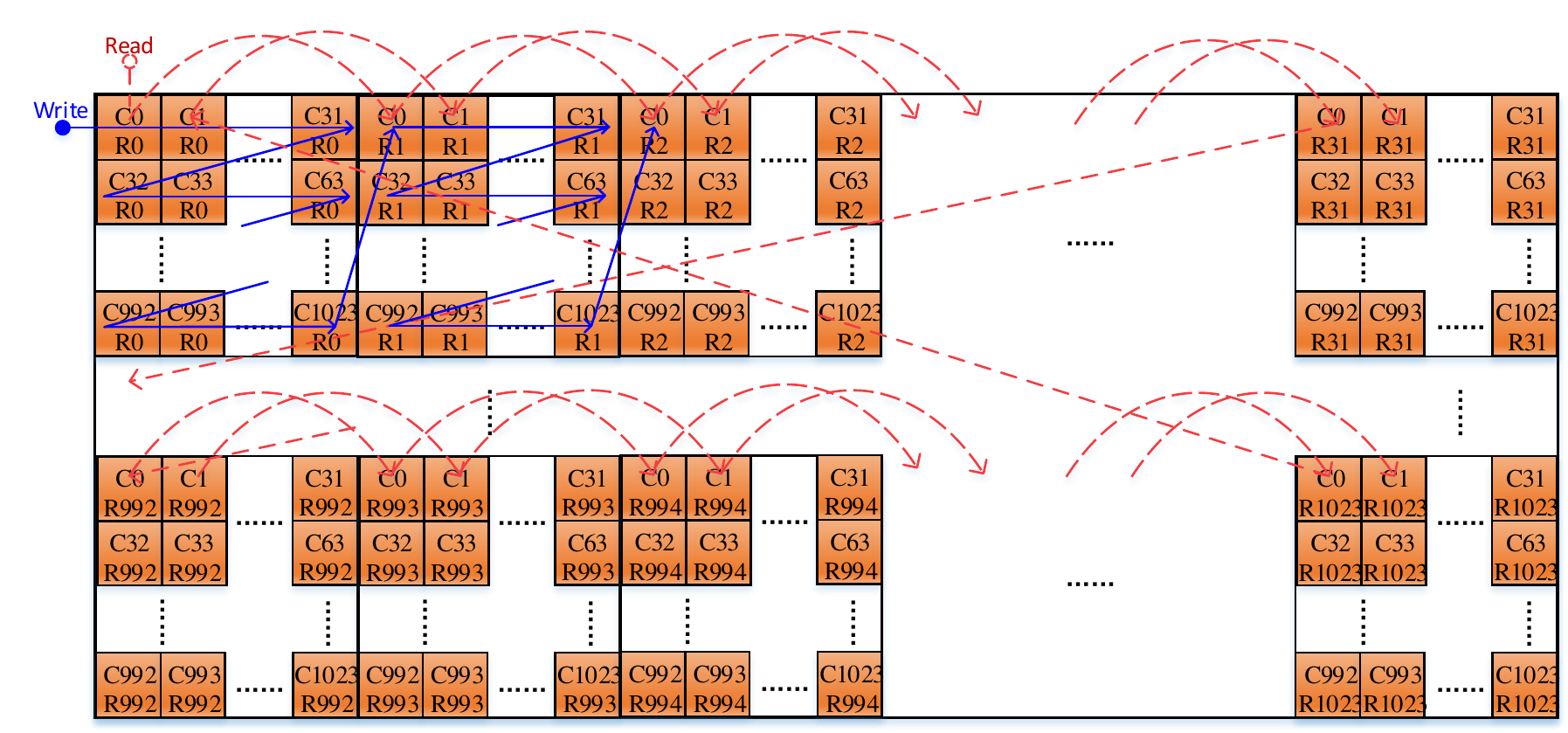}
	\caption{the High Effective Matrix Transposition Process}
	\label{fig_fast_matrix}
\end{figure}

The high effective matrix transposition method uses the matrix partitioning technology to balance the row span access times of the read operation and the write operation. This method can reduce the total row span access times and increase the data rates significantly. the main process of the high effective matrix transposition is indicated as Fig.\ref{fig_fast_matrix}.

In this case, each row of the matrix is transformed to a 32-dimension matrix. The row span access operation times of the high effective matrix transposition method is calculated as follow:

\begin{equation}
{T}_{row-span}={{T}_{write}}+{{T}_{read}}=32\times 1024+32\times 1024=65536
\label{eq12}
\end{equation}

This method reduces the row span access times obviously. This method is experimented with the DDR3-SDRAM to prove its improvement on the data rates of matrix transposition. The comparison experiment result of the two methods is indicated in TABLE \ref{table_1}.

\begin{table}[htp]
	\renewcommand{\arraystretch}{1.3}
	\caption{the High Effective Matrix Transposition Experiment Result}
	\label{table_1}
	\centering
	\begin{tabular}{c c c c c}
		\hline
		{} & Data Size & Matrix Format & Operation time & Average rate\\
		\hline
		The common method & 64Mb & $1024\times 1024$ & 8.959 us & 7.14Gbps\\ 
		The high effective method & 64Mb & $1024\times 1024$ & 3.242 us & 19.74Gbps \\
		\hline
	\end{tabular}
\end{table}

According to the experiment results, the high effective Matrix Transposition method can double the data rata of matrix transposition.

\section{Implementations and Results}

\begin{figure}[h]
	\centering
	\includegraphics[width=10cm, trim = 0 120 0 120 ]{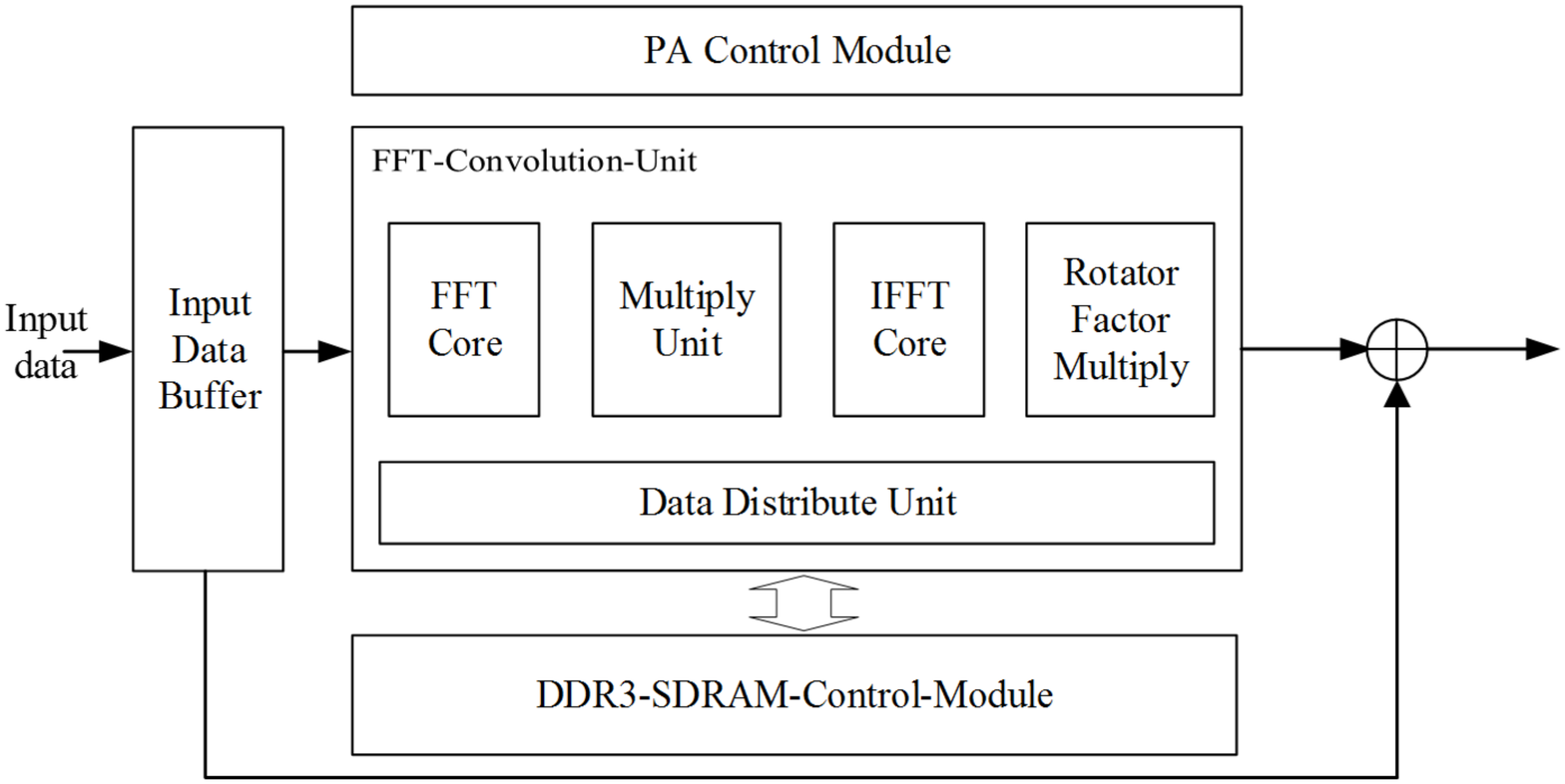}
	\caption{the overall structure of the PA scheme}
	\label{hardware-architecture}
\end{figure}

Our PA scheme is implemented on the ML605 Evaluation Kit. The kit is based on a Virtex-6 XC6VLX240T-1FFG1156 FPGA with 241,152 logic cells. The kit also contains a 512MB DDR3 SDRAM to support our scheme. The overall structure of our PA scheme is shown in Fig. \ref{hardware-architecture}.

The input data buffer is designed to store the input key and the Toeplitz random sequence. On output, It also converts the data to the floating-points for the FFT convolution. The PA Control module is the core controller of the PA module. It controls the FFT convolution unit to accomplish the PA computational tasks. The FFT convolution unit is the key unit in PA module. It contains five major parts. The FFT core is designed to calculate the FFT on each row of the matrix. Two FFT IP-Cores provided by Xilinx are used to meet the speed requirements. The length of the FFT IP-Core is set as 1024 in one million points PA scheme. Similarly, The IFFT core is designed to calculate the IFFT on each column or row of the matrix.  two FFT IP-Cores are used and their length also set as 1024. The multiply unit completes the computational task of the real-valued FFT accelerated optimizing. The rotator factory multiply unit point-wisely multiply rotation factors by the result of FFT/IFFT core to accomplish the two-dimensional FFT algorithm. The data distribute unit distributes the data for the calculation units and exchanges data with SDRAM controller. We simulated the scheme using the simulation tool Modelsim and verified the correctness of the scheme with the result on Matlab . Then the scheme was implemented on a ML605 Evaluation Kit and the result is accord with the simulation. The resource utilization of the PA scheme in hardware is shown in TABLE \ref{table_2}. According to the resource utilization above, there is enough spare resource for other modules to constitute the post-processing system on one chip with the PA module.

\begin{table}[htp]
	\renewcommand{\arraystretch}{1.3}
	\caption{the resource utilization of the PA scheme}
	\label{table_2}
	\centering
	\begin{tabular}{c c c c}
		\hline
		{} & The scheme used & XC6VLX240T Available
		 & Utilization rate\\
		\hline
		Number of LUTs & 37203 & 150720 & 24\% \\
		Number of 36K BRAMs & 164 & 416 & 39\% \\
		Number of 18K BRAMs & 66 & 832 & 7\% \\
		Number of DSP48E1s & 360 & 768 & 46\% \\
		\hline
	\end{tabular}
\end{table}

The comparison of several FPGA-based implementations of PA scheme is indicated in TABLE \ref{table_3}. The schemes in \cite{Zhang2012b},\cite{Constantin2017e},\cite{Yang2017c} are all based on the linear feedback shift register(LFSR) to calculate the Toeplitz matrix. This kind of scheme can be high speed and resource-saving when the compression ratio of PA is low and fixed. When the compression ratio of PA is high or floated, the resource consumption of PA will rapidly increase. The FFT-based PA algorithm implementation isn’t affected by the compression ratio, so it’s easy to implement the wide-range variable compression ratio with our PA scheme. Unlike the LFSR-based PA algorithm, the processing speed of the FFT-based PA algorithm is mainly limited by the memory transfer rate instead of the resource of FPGA. The processing speed of our PA scheme can reach 116Mbps and it can increase sharply with faster memory (e.g. DDR4-DRAM).

\begin{table}[htp]
	\renewcommand{\arraystretch}{2}
	\caption{the comparison of several FPGA-based PA schemes}
	\label{table_3}
	\centering
	\begin{tabular}{c c c c c}
		\hline
		{} & This Work & Yang et al.\cite{Yang2017c}
		& Constantin et al.\cite{Constantin2017e} & Zhang et al. \cite{Zhang2012b}\\
		\hline
		Devices & Xilinx Virtex-6 & Xilinx Virtex-7 & Xilinx Virtex-6 & Cyclone III\\
		Length of the final key & 0-1,000,000 & 0-1,000,000 & 0-995,328 & 76,800\\
		LUTs & 37,203 & 26,571 & 15,604 & 1,902\\
		BRAM & 7,092kb & 100kb & 0kb & 656kb\\
		Clock frequency & 200MHz & 100MHz & 125MHz & 40MHz\\
		Max. processing speed & 116Mbps & 64Mbps & 41Mbps & 0.7Mbps\\
		\hline
	\end{tabular}
\end{table}

\section{Conclusions and Outlook}

This paper provides a high-speed PA hardware scheme and its implementation in FPGA based on the FFT. 
The verification is accomplished on the Virtex-6 FPGA and the processing speed of the scheme can reach 116Mbps. Compared with other work, the proposed PA scheme supports wide-range and variable compression ratio and can reach faster processing speed with faster memory . The optimizing proposed in this paper can also improve the FFT-based PA algorithm on other platforms, such as CPU and GPU. In the future, we will research on the relationship between the precision of FFT and the safety of PA and try to replace the floating-point FFT with fixed-point FFT in PA algorithm to reduce the resource consumption.
\section*{Acknowledgements}


\bibliographystyle{IEEEtran}
\bibliography{Newlibrary}

\begin{thebibliography}{10}
\providecommand{\url}[1]{#1}
\csname url@samestyle\endcsname
\providecommand{\newblock}{\relax}
\providecommand{\bibinfo}[2]{#2}
\providecommand{\BIBentrySTDinterwordspacing}{\spaceskip=0pt\relax}
\providecommand{\BIBentryALTinterwordstretchfactor}{4}
\providecommand{\BIBentryALTinterwordspacing}{\spaceskip=\fontdimen2\font plus
\BIBentryALTinterwordstretchfactor\fontdimen3\font minus
  \fontdimen4\font\relax}
\providecommand{\BIBforeignlanguage}[2]{{%
\expandafter\ifx\csname l@#1\endcsname\relax
\typeout{** WARNING: IEEEtran.bst: No hyphenation pattern has been}%
\typeout{** loaded for the language `#1'. Using the pattern for}%
\typeout{** the default language instead.}%
\else
\language=\csname l@#1\endcsname
\fi
#2}}
\providecommand{\BIBdecl}{\relax}
\BIBdecl

\bibitem{Bennett2014f}
C.~H. Bennett and G.~Brassard, ``{Quantum cryptography: Public key distribution
  and coin tossing},'' \emph{Theoretical Computer Science}, vol. 560, no.~P1,
  pp. 7--11, 2014.

\bibitem{Stucki2005c}
D.~Stucki, N.~Brunner, N.~Gisin, V.~Scarani, and H.~Zbinden, ``{Fast and simple
  one-way quantum key distribution},'' \emph{Applied Physics Letters}, vol.~87,
  no.~19, pp. 1--3, 2005.

\bibitem{Scarani2004c}
V.~Scarani, A.~Ac{\'{i}}n, G.~Ribordy, and N.~Gisin, ``{Quantum Cryptography
  Protocols Robust against Photon Number Splitting Attacks for Weak Laser Pulse
  Implementations},'' \emph{Physical Review Letters}, vol.~92, no.~5, p.~4,
  2004.

\bibitem{Inoue2002c}
K.~Inoue, E.~Waks, and Y.~Yamamoto, ``{Differential phase shift quantum key
  distribution},'' \emph{Physical Review Letters}, vol.~89, no.~3, pp.
  379\,021--379\,023, 2002.

\bibitem{BruB1998c}
D.~Bru{\ss}, ``{Optimal eavesdropping in quantum cryptography with six
  states},'' \emph{Physical Review Letters}, vol.~81, no.~14, pp. 3018--3021,
  1998.

\bibitem{Bennett1992e}
C.~H. Bennett, ``{Quantum cryptography using any two nonorthogonal states},''
  \emph{Physical Review Letters}, vol.~68, no.~21, pp. 3121--3124, 1992.

\bibitem{Cerf2001c}
N.~J. Cerf, M.~L{\'{e}}vy, and G.~{Van Assche}, ``{Quantum distribution of
  Gaussian keys using squeezed states},'' \emph{Physical Review A. Atomic,
  Molecular, and Optical Physics}, vol.~63, no.~5, pp. 523\,111--523\,115,
  2001.

\bibitem{Grosshans2002f}
F.~Grosshans and P.~Grangier, ``{Continuous variable quantum cryptography using
  coherent states},'' \emph{Physical review letters}, pp. 1--6, 2002.

\bibitem{Pirandola2008c}
S.~Pirandola, S.~Mancini, S.~Lloyd, and S.~L. Braunstein,
  ``{Continuous-variable quantum cryptography using two-way quantum
  communication},'' \emph{Nature Physics}, vol.~4, no.~9, pp. 726--730, 2008.

\bibitem{SUN2012c}
M.~SUN, X.~PENG, Y.~SHEN, and H.~GUO, ``{Security of a New Two-Way
  Continuous-Variable Quantum Key Distribution Protocol},'' \emph{International
  Journal of Quantum Information}, vol.~10, no.~05, p. 1250059, 2012.

\bibitem{Wang2012b}
S.~Wang, W.~Chen, J.-F. Guo, Z.-Q. Yin, H.-W. Li, Z.~Zhou, G.-C. Guo, and Z.-F.
  Han, ``{2 GHz clock quantum key distribution over 260 km of standard telecom
  fiber},'' \emph{Optics Letters}, vol.~37, no.~6, p. 1008, 2012.

\bibitem{Muller2007b}
A.~Muller, J.~Breguet, and N.~Gisin, ``{Experimental Demonstration of Quantum
  Cryptography Using Polarized Photons in Optical Fibre over More than 1 km},''
  \emph{Europhysics Letters}, vol.~23, no.~6, pp. 383--388, 2007.

\bibitem{Townsend1994b}
P.~D. Townsend and I.~Thompson, ``{A quantum key distribution channel based on
  optical fibre},'' \emph{Journal of Modern Optics}, vol.~41, no.~12, pp.
  2425--2433, 1994.

\bibitem{Peng2007b}
C.~Z. Peng, J.~Zhang, D.~Yang, W.~B. Gao, H.~X. Ma, H.~Yin, H.~P. Zeng,
  T.~Yang, X.~B. Wang, and J.~W. Pan, ``{Experimental long-distance decoy-state
  quantum key distribution based on polarization encoding},'' \emph{Physical
  Review Letters}, vol.~98, no.~1, 2007.

\bibitem{Bennett1995d}
C.~H. Bennett, G.~Brassard, C.~Crkpeau, U.~M. Maurer, and S.~Member,
  ``{Generalized privacy amplification},'' \emph{Information Theory, IEEE
  Transactions on}, vol.~41, no.~6, pp. 1915--1923, 1995.

\bibitem{Bennett1988d}
C.~H. Bennett, G.~Brassard, and J.-M. Robert, ``{Privacy Amplification by
  Public Discussion},'' \emph{SIAM Journal on Computing}, vol.~17, no.~2, pp.
  210--229, 1988.

\bibitem{Impagliazzo1989c}
R.~Impagliazzo, L.~a. Levint, and M.~Luby, ``{Pseudo-random generation from
  one-way functions ( Extended Abstract )},'' \emph{STOC '89 Proceedings of the
  twenty-first annual ACM symposium on Theory of computing}, pp. 12--24, 1989.

\bibitem{Wegman1981b}
M.~N. Wegman and J.~L. Carter, ``{New hash functions and their use in
  authentication and set equality},'' \emph{Journal of Computer and System
  Sciences}, vol.~22, no.~3, pp. 265--279, 1981.

\bibitem{Carter1979b}
J.~L.~J. Carter and M.~M. N.~M. Wegman, ``{Classes of Hash Functions},''
  \emph{Journal of computer and system sciences}, vol.~18, pp. 143--154, 1979.

\bibitem{Cai2009c}
R.~Y. Cai and V.~Scarani, ``{Finite-key analysis for practical implementations
  of quantum key distribution},'' \emph{New Journal of Physics}, vol.~11, pp.
  1--17, 2009.

\bibitem{Zhang2014c}
C.~M. Zhang, M.~Li, J.~Z. Huang, H.~W. Li, F.~Y. Li, C.~Wang, Z.~Q. Yin,
  W.~Chen, Z.~F. Han, P.~Treeviriyanupab, and K.~Sripimanwat, ``{Fast
  implementation of length-adaptive privacy amplification in quantum key
  distribution},'' \emph{Chinese Physics B}, vol.~23, no.~9, pp. 1--6, 2014.

\bibitem{Krawczyk1994b}
H.~Krawczyk, ``{LFSR-based Hashing and Authentication},'' \emph{Advances in
  Cryptology — CRYPTO '94}, vol. 10598, pp. 129--139, 1994.

\bibitem{Zhang2012b}
H.~F. Zhang, J.~Wang, K.~Cui, C.~L. Luo, S.~Z. Lin, L.~Zhou, H.~Liang, T.~Y.
  Chen, K.~Chen, and J.~W. Pan, ``{A real-time QKD system based on FPGA},''
  \emph{Journal of Lightwave Technology}, vol.~30, no.~20, pp. 3226--3234,
  2012.

\bibitem{Constantin2017e}
J.~Constantin, R.~Houlmann, N.~Preyss, N.~Walenta, H.~Zbinden, P.~Junod, and
  A.~Burg, ``{An FPGA-Based 4 Mbps Secret Key Distillation Engine for Quantum
  Key Distribution Systems},'' \emph{Journal of Signal Processing Systems},
  vol.~86, no.~1, pp. 1--15, 2017.

\bibitem{Yang2017c}
S.~S. Yang, Z.~L. Bai, X.~Y. Wang, and Y.~M. Li, ``{FPGA-Based Implementation
  of Size-Adaptive Privacy Amplification in Quantum Key Distribution},''
  \emph{IEEE Photonics Journal}, vol.~9, no.~6, 2017.

\bibitem{Liu2016d}
B.~Liu, B.~Zhao, W.~Yu, and C.~Wu, ``{FiT-PA: Fixed scale FFT based privacy
  amplification algorithm for quantum key distribution},'' \emph{Journal of
  Internet Technology}, vol.~17, no.~2, pp. 309--320, 2016.

\bibitem{XiangyuWangYichenZhangSongYu2016e}
{Xiangyu Wang, Yichen Zhang, Song Yu} and H.~Guo, ``{High-Speed Implementation
  of Privacy Amplification in Quantum Key Distribution},'' vol.~xx, no.~xx, pp.
  1--10, 2016.

\bibitem{Scarani2008c}
V.~Scarani and R.~Renner, ``{Quantum cryptography with finite resources:
  Unconditional security bound for discrete-variable protocols with one-way
  postprocessing},'' \emph{Physical Review Letters}, vol. 100, no.~20, pp.
  1--4, 2008.

\bibitem{Urumaru2013a}
T.~Urumaru, M.~Hayashi, and S.~Member, ``{Dual Universality of Hash Functions
  and Its Applications to Quantum Cryptography},'' vol.~59, no.~7, pp.
  4700--4717, 2013.

\bibitem{LernerParallel2018Spec}
B.~Lerner, ``Parallel implementation of fixed-point ffts on tigersharc
  processors.''

\bibitem{Sorensen1987a}
H.~V. Sorensen, D.~L. Jones, M.~T. Heideman, and C.~S. Burrus, ``{Real-Valued
  Fast Fourier Transform Algorithms},'' \emph{IEEE Transactions on Acoustics,
  Speech, and Signal Processing}, vol.~35, no.~6, pp. 849--863, 1987.

\bibitem{Qinwen2017}
W.~U. Qinwen, ``{High Efficiency Matrix Transposition Method Based on FPGA and
  DDR},'' 2017.

\end{thebibliography}



\end{document}